**Elastic, electronic, and optical properties of $Sb_2S_3$ and $Sb_2Se_3$ compounds: ab initio calculation**

H. Koc[1*], Amirullah M. Mamedov[2], E. Deligoz[3], and H. Ozisik[4]

[1] Department of Physics, Siirt University, 56100 Siirt, Turkey
[2]Nanotechnology Research Center (NANOTAM), Bilkent University, 06800 Bilkent, Ankara, Turkey
[3]Department of Physics, Aksaray University, 68100 Aksaray, Turkey
[4]Aksaray University, Department of Computer and Instructional Technologies Teaching, 68100 Aksaray, Turkey

**Abstract**

We have performed a first principles study of structural, mechanical, electronic, and optical properties of orthorhombic $Sb_2S_3$ and $Sb_2Se_3$ compounds using the density functional theory within the local density approximation. The lattice parameters, bulk modulus, and its pressure derivatives of these compounds have been obtained. The second-order elastic constants have been calculated, and the other related quantities such as the Young's modulus, shear modulus, Poisson's ratio, anisotropy factor, sound velocities, Debye temperature, and hardness have also been estimated in the present work. The linear photon-energy dependent dielectric functions and some optical properties such as the energy-loss function, the effective number of valance electrons and the effective optical dielectric constant are calculated. Our structural estimation and some other results are in agreement with the available experimental and theoretical data.

## 1. Introduction

$Sb_2S_3$ and $Sb_2Se_3$, a member of compounds with the general formula $A_2^V B_3^{VI}$ ($A$=Bi, Sb and $B$=S, Se), are layer-structured semiconductors with orthorhombic crystal structure (space group Pnma; No:62), in which each Sb-atom and each Se(S)-atom is bound to three atoms of the opposite kind that are then held together in the crystal by weak secondary bond [1, 2]. In the last few years, $Sb_2Se_3$ has received a great deal of attention due to its switching effects [3] and its excellent photovoltaic properties and high thermoelectric power [4], which make it possess promising applications in solar selective and decorative coating, optical and thermoelectric cooling devices [5]. On the other hand, $Sb_2S_3$ has attracted attention for its applications as a target material for TV systems [6,7], as well as in microwave [8], switching [9], and optoelectronic devices [10-12].

The crystal structure of $Sb_2S_3$ and $Sb_2Se_3$ are shown in Fig. 1. The positions corresponding to the orthorhombic $Sb_2S_3$ and $Sb_2Se_3$ have been obtained from experimental data [13-15]. The atomic positions are given in Table 1. These crystals have four $Sb_2B_3$ (B=S, Se) molecules (20 atoms) in unit cell. Therefore, these compounds have a complex structure with 112 valance electrons per unit cell.

In the past, some detailed works [15-17] have been carried out on the structural and electronic properties of these compounds. The valance electron density, the electron band structure, and the corresponding electronic density-of-states (DOS) of $A_2B_3$ (A=Bi, Sb and B=S, Se) compounds using the density functional theory were studied by Caracas et al [15]. Ben Nasr et al [16] computed the electronic band structure, density of states, charge density and optical properties, such as the dielectric function, reflectivity spectra, refractive index and the loss function using the full potential linearized augmented plane waves (FP-LAPW) method as implemented in the Wien2k code for $Sb_2S_3$. Kuganathan et al [17] used density functional methods as embedded in the SIESTA code, to test the proposed model theoretically and investigate the perturbations on the molecular and electronic structure of the crystal and the SWNT (single walled carbon nanotubes) and the energy of formation of the $Sb_2Se_3$ SWNT composite.

As far as we know, no ab initio general potential calculations of the elastic constants, Young's modulus, shear modulus, Poisson's ratio, anisotropy factor, sound velocities, Debye temperature, and optical properties such as the energy-loss function, the effective number of valance electrons and the effective optical dielectric constant along y- and z- axes of the $Sb_2S_3$ and $Sb_2Se_3$ have been reported in detail. In the present work, we have investigated the structural, electronic, mechanical, and photon energy-dependent optical properties of the $Sb_2S_3$ and $Sb_2Se_3$ crystals. The method of calculation is given in Section 2; the results are discussed in Section 3. Finally, the summary and conclusion are given in Section 4.

## 2. Method of calculation

Simulations of $Sb_2S_3$ and $Sb_2Se_3$ compounds were conducted, using two different Quantum Mechanical (QM) DFT programs. The first, freely accessible code, SIESTA combines norm conserving pseudopotentials with the local basis functions. The calculations were performed using the density functional formalism and local density approximation (LDA) [18] through the Ceperley and Alder functional [19] as parameterized by Perdew and Zunger [20] for the exchange-correlation energy in the SIESTA code [21, 22]. This code calculates the total energies and atomic forces using a linear combination of atomic orbitals as the basis set. The basis set is based on the finite range pseudoatomic orbitals (PAOs) of the Sankey_Niklewsky type [23], generalized to include multiple-zeta decays.

The interactions between electrons and core ions are simulated with separable Troullier-Martins [24] norm-conserving pseudopotentials. We have generated atomic pseudopotentials separately for atoms, Sb, S and Se by using the $5s^2 5p^3$, $3s^2 3p^4$ and $4s^2 4p^4$ configurations, respectively. The cut-off radii for present atomic pseudopotentials are taken as s: 1.63 au, p: 1.76 au, 1.94 au for the d and f channels of S, s: 1.94 au, p: 2.14 au, d: 1.94 au. f: 2.49 of Se and 2.35 for the s, p, d and f channels of Sb.

Siesta calculates the self-consistent potential on a grid in real space. The fineness of this grid is determined in terms of an energy cut-off $E_c$ in analogy to the energy cut-off when the basis set involves plane waves. Here by using a double-zeta plus polarization (DZP) orbitals basis and the cut-off energies between 100 and 450 $Ry$ with various basis sets, we found an optimal value of around 350 $Ry$ for $Sb_2S_3$ and $Sb_2Se_3$. For the final computations, 256 k-points for $Sb_2S_3$ and $Sb_2Se_3$ were enough to obtain the converged total energies ΔE to about 1meV/atoms.

The second, commercially available (VASP) [25-28] code, employs plane wave basis functions. The calculations were performed with this code and reported here also use the LDA. The electron-ion interaction was considered in the form of the projector-augmented-wave (PAW) method with plane wave up to energy of 450 eV [28, 29]. This cut-off was found to be adequate for studying the structural and elastic properties. The 8x11x8 Monkhorst and Pack [30] grid of k-points have been used for these compounds.

## 3. Results and discussion

### 3.1 Structural properties

All physical properties are related to the total energy. For instance, the equilibrium lattice constant of a crystal is the lattice constant that minimizes the total energy. If the total energy is calculated, any physical property related to the total energy can be determined.

For $Sb_2S_3$ and $Sb_2Se_3$, structures which are orthorhombic are considered. The equilibrium lattice parameters, the bulk modulus, and its pressure derivative have been computed minimizing the crystal's total energy calculated for the different values of lattice constant by means of Murnaghan's equation of states (eos) [31] for SIESTA calculations. We have fully relaxed the cell volume and the ionic positions of atoms in reciprocal coordinates which is supported by VASP code [25-28] for all considered compounds. In all calculations, we have used these relaxed parameters for VASP calculations. The results for SIESTA and VASP calculations are shown in Table 2 along with the experimental and theoretical values. The lattice parameters obtained using SIESTA and VASP for $Sb_2S_3$ and $Sb_2Se_3$ are in a good agreement with the experimental and theoretical values. It is seen that the lattice parameter values of SIESTA compared to experimental and theoretical values are better than values obtained by VASP. In all our further calculations, we have used the computed lattice parameters. In the present case, the calculated bulk moduli of SIESTA for $Sb_2S_3$ and $Sb_2Se_3$ are 73.64 and 64.78 GPa, respectively. The bulk modulus for $Sb_2S_3$ are higher (about 2.7%) than the other theoretical result given in Ref. [16]. This small difference may stem from the different density- functional-based electronic structure methods.

## 3.2. Elastic properties

The elastic constant of solids provides a link between the mechanical and dynamical behavior of crystals, and give important information concerning the nature of the forces operating in solids. In particular, they provide information on the stability and stiffness of materials, and their ab initio calculation requires precise methods. Since the forces and the elastic constants are functions of the first-order and second-order derivatives of the potentials, their calculation will provide a further check on the accuracy of the calculation of forces in solids. They also provide valuable data for developing inter atomic potentials [37-40].

Here, to compute the elastic constants $(C_{ij})$, we have used the "volume-conserving" technique [41] for SIESTA calculations. We have also derived the elastic constants from the strain-stress relationship [42] for VASP calculations. The elastic constants for SIESTA and VASP calculations are given in Table 3. The elastic constant values of SIESTA are, generally, in accord with the elastic constant values of VASP. But, the calculated value of $C_{44}$ for VASP is lower than the result of SIESTA, whereas $C_{22}$ and $C_{13}$ values for VASP is higher than the SIESTA results. So, further study is necessary to solve the discrepancy. These differences many originate from the different density-functional based electronic structure methods. Unfortunately, there are no theoretical results for comparing with the present work. Then, our results can serve as a prediction for future investigations.

Nine independent strains are necessary to compute the elastic constants of orthorhombic $Sb_2S_3$ and $Sb_2Se_3$ compounds. Mechanical stability leads to restrictions on the elastic constants, which for orthorhombic crystals [41, 43, 44] are

$$(C_{11}+C_{22}-2C_{12})>0,\ (C_{11}+C_{33}-2C_{13})>0,$$
$$(C_{22}+C_{33}-2C_{23})>0,\ C_{11}>0,\ C_{22}>0,$$
$$C_{33}>0,\ C_{44}>0,\ C_{55}>0,\ C_{66}>0, \quad (1)$$
$$(C_{11}+C_{22}+C_{33}+2C_{12}+2C_{13}+2C_{23})>0.$$

The present elastic constants in Table 3 obey these stability conditions for orthorhombic $Sb_2S_3$ and $Sb_2Se_3$.

The elastic constants $C_{11}$, $C_{22}$, and $C_{33}$ measure the a-, b-, and c- direction resistance to linear compression, respectively. The $C_{22}$ for SIESTA calculations is lower than the $C_{11}$ and $C_{33}$ while the $C_{33}$ for VASP calculations of $Sb_2S_3$ is lower than the $C_{11}$ and $C_{22}$. The calculated $C_{33}$ of both code for $Sb_2Se_3$ are lower than the $C_{11}$ and $C_{22}$. Thus, $Sb_2S_3$ compound is more compressible along b axis for SIESTA calculations and c axis for VASP calculations while $Sb_2Se_3$ compound is more compressible along c axis for both codes.

It is known that, the elastic constant $C_{44}$ is the most important parameter indirectly governing the indentation hardness of a material. The large $C_{44}$ means a strong ability of resisting the monoclinic shear distortion in (100) plane, and the $C_{66}$ relates to the resistance to shear in the <110> direction. In the present case, $C_{44}$ and $C_{66}$ for both codes of $Sb_2S_3$ is higher than $Sb_2Se_3$ compound.

A problem arises when single crystal samples are not available, since it is then not possible to measure the individual elastic constants. Instead, the polycrystalline bulk modulus ($B$) and shear modulus ($G$) may be determined. There are two approximation methods to calculate the polycrystalline modulus, namely, the Voigt method [45] and the Reuss method [46]. For specific cases of orthorhombic lattices, the Reuss shear modulus ($G_R$) and the Voigt shear modulus ($G_V$) are

$$G_R = 15\left\{\begin{matrix}[C_{11}(C_{22}+C_{33}+C_{23})+C_{22}(C_{33}+C_{13})+C_{33}C_{12}-C_{12}(C_{23}+C_{12})-C_{13}(C_{12}+C_{13}) \\ -C_{23}(C_{13}+C_{23})]/\Delta+3[(1/C_{44})+(1/C_{55})+(1/C_{66})]\end{matrix}\right\}^{-1} \quad (2)$$

and

$$G_V = \frac{1}{15}(C_{11} + C_{22} + C_{33} - C_{12} - C_{13} - C_{23} + \frac{1}{5}(C_{44} + C_{55} + C_{66}), \tag{3}$$

and the Reuss bulk modulus ( $B_R$ ) and Voight bulk modulus ($B_V$) are defined as

$$B_R = \Delta\begin{bmatrix} C_{11}(C_{22} + C_{33} - 2C_{23}) + C_{22}(C_{33} - 2C_{13}) - 2C_{33}C_{12} + C_{12}(2C_{23} - C_{12}) + C_{13}(2C_{12} - C_{13}) \\ + C_{23}(2C_{13} - C_{23}) \end{bmatrix}^{-1} \tag{4}$$

and

$$B_V = \frac{1}{9}(C_{11} + C_{22} + C_{33}) + \frac{2}{9}(C_{12} + C_{13} + C_{23}) \tag{5}$$

In Eq. (2) and (4), the $\Delta = C_{13}(C_{12}C_{23} - C_{13}C_{22}) + C_{23}(C_{12}C_{13} - C_{23}C_{11}) + C_{33}(C_{11}C_{22} - C_{12}^2)$ is elastic compliance constant. Using energy considerations Hill [47] proved that the Voigt and Reuss equations represent upper and lower limits of the true polycrystalline constants, and recommended that a practical estimate of the bulk and shear moduli were the arithmetic means of the extremes. Hence, the elastic moduli of the polycrystalline material can be approximated by Hill's average and for shear moduli it is

$$G = \frac{1}{2}(G_R + G_V) \tag{6}$$

and for bulk moduli it is

$$B = \frac{1}{2}(B_R + B_V) \tag{7}$$

The Young's modulus, $E$ , and Poisson's ratio, $v$ , for an isotropic material are given by

$$E = \frac{9BG}{3B + G} \tag{8}$$

$$v = \frac{3B - 2G}{2(3B + G)}, \tag{9}$$

respectively. [48,49]. Using the relations given above the calculated bulk modulus, shear modulus, Young's modulus, and Poisson's ratio of both codes for $Sb_2S_3$ and $Sb_2Se_3$ are give Table 4.

It is known that isotropic shear modulus and bulk modulus are a measure of the hardness of a solid. The bulk modulus is a measure of resistance to volume change by an applied pressure, whereas the shear modulus is a measure of resistance to reversible deformations upon shear stress [50]. Therefore, isotropic shear modulus is better predictor of hardness than the bulk modulus. The isotropic shear modulus, a measurement of resistance to shape change, is more pertinent to hardness and the larger shear modulus is mainly due to its larger $C_{44}$. The calculated isotropic shear modulus and bulk modulus of SIESTA (VASP) are 47.49 (47.11), 75.10 (80.25) GPa and 33.05 (41.51), 58.74 (70.52) GPa for $Sb_2S_3$ and $Sb_2Se_3$, respectively. The values of the bulk moduli indicate that $Sb_2S_3$ is less compressible material than $Sb_2Se_3$ compound. The calculated shear modulus for $Sb_2S_3$ is higher than $Sb_2Se_3$ compound.

According to the criterion in refs. [50, 51], a material is brittle (ductile) if the $B/G$ ratio is less (high) than 1.75. The value of the $B/G$ of SIESTA calculations is lower and higher than 1.75 for $Sb_2S_3$ and $Sb_2Se_3$, respectively. Hence, $Sb_2S_3$ behave in a brittle manner while $Sb_2Se_3$ behave in a ductile manner. For VASP calculations, the value of the $B/G$ of both compounds are lower than 1.75. Hence, both compounds behave in a brittle. So, further study is necessary to solve the discrepancy.

Young's modulus is defined as the ratio of stress and strain, and used to provide a measure of the stiffness of the solid. The material is stiffer if the value of Young's modulus is high. In this context, due to the higher value of Young's modulus (117.66 GPa for SIESTA and 118.20 for VASP) $Sb_2S_3$ compound is relatively stiffer than $Sb_2Se_3$ (83.49 GPa for SIESTA and 104.52 for VASP). If the value of E, which has an impact on the ductile, increases, the covalent nature of the material also increases. From Table 4, one can see that E increases as one moves from $Sb_2Se_3$ to $Sb_2S_3$.

The value of the Poisson's ratio is indicative of the degree of directionality of the covalent bonds. The value of the Poisson's ratio is small ($\upsilon$ =0.1) for covalent materials, whereas for ionic materials a typical value of $\upsilon$ is 0.25 [52]. The calculated Poisson's ratios of SIESTA and VASP are about 0.24, 0.25 and 0.26, 0.25 for $Sb_2S_3$ and $Sb_2Se_3$, respectively. Therefore, the ionic contribution to inter atomic bonding for these compounds is dominant. The $\upsilon$=0.25 and 0.5 are the lower and upper limits, respectively, for central force solids [53]. Our $\upsilon$ values are close to the value of 0.25 indicating inter atomic forces are weightlessly central forces in $Sb_2S_3$ and $Sb_2Se_3$.

Many low symmetry crystals exhibit a high degree of elastic anisotropy [54]. The shear anisotropic factors on different crystallographic planes provide a measure of the degree of anisotropy in atomic bonding in different planes. The shear anisotropic factors are given by

$$A_1 = \frac{4C_{44}}{C_{11} + C_{33} - 2C_{13}} \text{ for the \{100\} plane} \qquad (10)$$

$$A_2 = \frac{4C_{55}}{C_{22} + C_{33} - 2C_{23}} \text{ for the \{010\}plane} \tag{11}$$

$$A_3 = \frac{4C_{66}}{C_{11} + C_{22} - 2C_{12}} \text{ for the \{001\}plane} \tag{12}$$

The calculated $A_1, A_2$ and $A_3$ of both code for $Sb_2S_3$ and $Sb_2Se_3$ are given in Table 5. A value of unity means that the crystal exhibits isotropic properties while values other than unity represent varying degrees of anisotropy. From Table 5, it can be seen that $Sb_2S_3$ and $Sb_2Se_3$ exhibit larger anisotropy in the {100} and {010}planes and these compounds exhibits almost isotropic properties for the {001} plane according to other planes. Another way of measuring the elastic anisotropy is given by the percentage of anisotropy in the compression and shear [52, 53, 55].

$$A_{comp} = \frac{B_V - B_R}{B_V + B_R} x100 \tag{13}$$

$$A_{shear} = \frac{G_V - G_R}{G_V + G_R} x100 \tag{14}$$

For crystals, these values can range from zero (isotropic) to 100% representing the maximum anisotropy. The percentage anisotropy values have been computed for $Sb_2S_3$ and $Sb_2Se_3$, and are shown in Table 5. It can be also seen that the anisotropy in compression is small and the anisotropy in shear is high. $Sb_2S_3$ compound exhibit relatively high shear and bulk anisotropies among these compounds.

The Debye temperature is known as an important fundamental parameter closely related to many physical properties such as specific heat and melting temperature. At low temperatures the vibrational excitations arise solely from acoustic vibrations. Hence, at low temperatures the Debye temperature calculated from elastic constants is the same as that determined from specific heat measurements. We have calculated the Debye temperature, $\theta_D$, from the elastic constants data using the average sound velocity, $v_m$, by the following common relation given in Ref. [56]

$$\theta_D = \frac{\hbar}{k}\left[\frac{3n}{4\pi}\left(\frac{N_A \rho}{M}\right)\right]^{1/3} v_m, \tag{15}$$

where $\hbar$ is Planck's constants, $k$ is Boltzmann's constant, $N_A$ is Avogadro's number, n is the number of atoms per formula unit, $M$ is the molecular mass per formula unit, $\rho(=M/V)$ is the density, and $v_m$ is given [57] as

$$v_m = \left[ \frac{1}{3} \left( \frac{2}{v_t^3} + \frac{1}{v_l^3} \right) \right]^{-1/3}, \tag{16}$$

where $v_l$ and $v_t$, are the longitudinal and transverse elastic wave velocities, respectively, which are obtained from Navier's equation [58]

$$v_l = \sqrt{\frac{3B + 4G}{3\rho}}, \tag{17}$$

and

$$v_t = \sqrt{\frac{G}{\rho}} \tag{18}$$

The calculated values of the longitudinal, transverse, average sound velocities and density in the present formalism for SIESTA and VASP calculations are shown in Table 6 along with the Debye temperature. For materials, usually, the higher Debye temperature, the larger microhardness. The calculated Debye temperature for $Sb_2S_3$ is higher than $Sb_2Se_3$. Unfortunately, there are no theoretical and experimental results to compare with the calculated $v_l$, $v_t$, $v_m$ , and $\theta_D$ values.

Recently, Chen et al. [59] proposed new theoretical model to predict the hardness of polycrystalline materials based on the squared Pugh's modulus ratio (*k=G/B*) and the shear modulus (*G*) as below:

$$H_v = 2\left(k^2 G\right)^{0.585} - 3 \tag{19}$$

We have used to predict the Vicker hardness of the considered compounds by using Eq. 19 and the results are listed in Table 6. The results indicate that the hardness of $Sb_2S_3$ is higher than $Sb_2Se_3$, and they show soft character.

## 3.3. Electronic properties

For a better understanding of the electronic and optical properties of $Sb_2S_3$ and $Sb_2Se_3$, the investigation of the electronic band structure would be useful. The electronic band structures of orthorhombic $Sb_2S_3$ and $Sb_2Se_3$ single crystals have been calculated along high symmetry directions in the first Brillouin zone (BZ) using the results of SIESTA calculations. The band structures were calculated along the special lines connecting the high-symmetry points S

(1/2,1/2,0), Y (0,1/2,0), $\Gamma$ (0,0,0), S(1/2,1/2,0), R (1/2,1/2,1/2) for $Sb_2S_3$ and $Sb_2Se_3$ in the k-space. The results of the calculation are shown in Fig. 2 for these single crystals.

The energy band structures calculated using LDA for $Sb_2S_3$ and $Sb_2Se_3$ are shown in Fig. 2. As can be seen in Fig. 2a, the $Sb_2S_3$ compound has an direct band gap semiconductor with the value 1.18 $eV$. The top of the valance band and the bottom of the conduction band positioned at the $\Gamma$ point of BZ. The estimates of the band gap of $Sb_2S_3$ are contradictory in the literature. The band gap values estimated for $Sb_2S_3$ vary from 1.56 $eV$ to 2.25 $eV$ (see Table 7). In conclusion, our band gap value obtained is different from experimental and theoretical values and the band gap has same character with given in Ref. [62-65, 16]. The present band and the density of states (DOS) profiles for $Sb_2S_3$ agree with the earlier work [16]

The calculated band structure of $Sb_2Se_3$ is given in Fig. 2b. As can be seen from the figure, the band gap has the different character with that of $Sb_2S_3$, that is, it is an indirect band gap semiconductor. The top of the valance band positioned at the nearly midway between $\Gamma$ and S point of BZ, the bottom of the conduction band is located at the nearly midway between the $\Gamma$ and Y point of BZ. The indirect and direct band gap values of $Sb_2Se_3$ compound are, 0.99 $eV$ and 1.07 $eV$, respectively. The band gap values estimated for $Sb_2Se_3$ vary from 1.56 $eV$ to 2.25 $eV$ (see Table 7). Our band gap value obtained is good agreement with experimental and theoretical values and the character of the band gap is different from that given in Ref. [66,67].

The total and partial densities of states of $Sb_2S_3$ and $Sb_2Se_3$ are illustrated in Fig. 3. As you can see, from this figure, the lowest valence bands occur between about -14 and -12 $eV$ are dominated by S 3s and Se 4s states while valence bands occur between about -10 and -7 $eV$ are dominated by Sb 5s states. The highest occupied valance bands are essentially dominated by S 3p and Se 4p states. The 5p states of Sb atoms are also contributing to the valance bands, but the values of densities of these states are so small compared to S 3p and Se 4p states. The lowest unoccupied conduction bands just above Fermi energy level is dominated by Sb 5p. The 3p (4p) states of S (Se) atoms are also contributing to the conduction bands, but the values of densities of these states are so small compared to Sb 5p states

Band structures of $Sb_2S_3$ and $Sb_2Se_3$ single crystals are compared, band structures of these crystals are highly resemble one another. Thus, on formation of the band structures of $Sb_2S_3$ and $Sb_2Se_3$ the 5s 5p orbitals of Sb atoms are more dominant than 3s3p and 4s4p orbitals of S and Se atoms. Finally, the band gap values obtained are less than the estimated experimental and theoretical results. For all crystal structures considered, the band gap values are underestimated than the experimental values. This state is caused from the exchange-correlation approximation of DFT.

3.4. Optical properties

It is well known that the effect of the electric field vector, **E**$(\omega)$, of the incoming light is to polarize the material. At the level of linear response, this polarization can be calculated using the following relation [69]:

$$P^i(\omega) = \chi_{ij}^{(1)}(-\omega,\omega).E^j(\omega), \qquad (20)$$

where $\chi_{ij}^{(1)}$ is the linear optical susceptibility tensor and it is given by [70]

$$\chi_{ij}^{(1)}(-\omega,\omega) = \frac{e^2}{\hbar\Omega}\sum_{nm\vec{k}} f_{nm}(\vec{k})\frac{r_{nm}^i(\vec{k})r_{mn}^i(\vec{k})}{\omega_{mn}(\vec{k})-\omega} = \frac{\varepsilon_{ij}(\omega)-\delta_{ij}}{4\pi} \qquad (21)$$

where $n,m$ denote energy bands, $f_{mn}(\vec{k}) \equiv f_m(\vec{k}) - f_n(\vec{k})$ is the Fermi occupation factor, $\Omega$ is the normalization volume. $\omega_{mn}(\vec{k}) \equiv \omega_m(\vec{k}) - \omega(\vec{k})$ are the frequency differences, $\hbar\omega_n(\vec{k})$ is the energy of band $n$ at wave vector **k.** The $\vec{r}_{nm}$ are the matrix elements of the position operator [70].

As can be seen from Eq. (21), the dielectric function $\varepsilon_{ij}(\omega) = 1 + 4\pi\chi_{ij}^{(1)}(-\omega,\omega)$ and the imaginary part of $\varepsilon_{ij}(\omega)$, $\varepsilon_2^{ij}(\omega)$, is given by

$$\varepsilon_2^{ij}(w) = \frac{e^2}{\hbar\pi}\sum_{nm}\int d\vec{k} f_{nm}(\vec{k})\frac{v_{nm}^i(\vec{k})v_{nm}^j(\vec{k})}{\omega_{mn}^2}\delta(\omega-\omega_{mn}(\vec{k})). \qquad (22)$$

The real part of $\varepsilon_{ij}(\omega)$,$\varepsilon_1^{ij}(\omega)$, can be obtained by using the Kramers-Kroning transformation [70]. Because the Kohn-Sham equations determine the ground state properties, the unoccupied conduction bands as calculated have no physical significance. If they are used as single-particle states in a calculation of optical properties for semiconductors, a band gap problem comes into included in calculations of response. In order to take into account self-energy effects, in the present work, we used the 'scissors approximation' [69].

In the present work, $\Delta$, the scissor shift to make the theoretical band gap match the experimental one, is 0.38 $eV$ and 0.11 $eV$ for $Sb_2S_3$ and $Sb_2Se_3$, respectively

The known sum rules [71] can be used to determine some quantitative parameters, particularly the effective number of the valence electrons per unit cell $N_{eff}$, as well as the effective optical dielectric constant $\varepsilon_{eff}$, which make a contribution to the optical constants of a crystal at the energy $E_0$. One can obtain an estimate of the distribution of oscillator strengths for both intraband and interband transitions by computing the $N_{eff}(E_0)$ defined according to

$$N_{eff}(E) = \frac{2m\varepsilon_0}{\pi\hbar^2 e^2 Na} \int_0^\infty \varepsilon_2(E) E dE, \tag{23}$$

Where $N_a$ is the density of atoms in a crystal, $e$ and $m$ are the charge and mass of the electron, respectively and $N_{eff}(E_0)$ is the effective number of electrons contributing to optical transitions below an energy of $E_0$.

Further information on the role of the core and semi-core bands may be obtained by computing the contribution which the various bands make to the static dielectric constant, $\varepsilon_0$. According to the Kramers-Kronig relations, one has

$$\varepsilon_0(E) - 1 = \frac{2}{\pi} \int_0^\infty \varepsilon_2(E) E^{-1} dE. \tag{24}$$

One can therefore define an 'effective' dielectric constant, which represents a different mean of the interband transitions from that represented by the sum rule, Eq. (24), according to the relation

$$\varepsilon_{eff}(E) - 1 = \frac{2}{\pi} \int_0^{E_0} \varepsilon_2(E) E^{-1} dE. \tag{25}$$

The physical meaning of $\varepsilon_{eff}$ is quite clear: $\varepsilon_{eff}$ is the effective optical dielectric constant governed by the interband transitions in the energy range from zero to $E_0$, i.e. by the polarizition of the electron shells.

In order to calculate the optical response by using the calculated band structure, we have chosen a photon-energy range of 0-25 $eV$ and have seen that a 0-17 $eV$ photon-energy range is sufficient for most optical functions.

The $Sb_2S_3$ and $Sb_2Se_3$ single crystals have an orthorhombic structure that is optically a biaxial system. For this reason, the linear dielectric tensor of the $Sb_2S_3$ and $Sb_2Se_3$ compounds have three independent components that are the diagonal elements of the linear dielectric tensor.

We first calculated the real and imaginary parts of the y- and z-components of the frequency-dependent linear dielectric function and these are shown in Fig. 4 and Fig. 5. The $\varepsilon_1^y$ behaves mainly as a classical oscillator. It vanishes (from positive to negative ) at about 3.48 $eV$ , 9.36 $eV$, 12.59 $eV$, and 20.69 $eV$, ( at the W, X, Y, and Z points in Fig. 4), whereas the other function $\varepsilon_1^z$ is equal to zero at about 3.57 $eV$, 9.46 $eV$, 13.10 $eV$ and 20.36 $eV$ (at the W, X, Y, Z points in Fig. 4) for $Sb_2S_3$ compound. The $\varepsilon_1^y$ is equal to zero at about 2.76 $eV$, 8.77 $eV$, 12.63 $eV$, and 20.04 $eV$, ( at the W, X, Y, and Z points in Fig. 5), whereas the other function $\varepsilon_1^z$ is equal to zero at about 2.95 $eV$, 8.91 $eV$, 13.06 $eV$ and 19.89 $eV$ (at the W, X, Y, Z points in Fig. 5) for $Sb_2Se_3$ compound. The peaks of the $\varepsilon_2^y$ and $\varepsilon_2^z$ correspond to the optical transitions from the valence band to the conduction band and are in agreement with the previous results. The maximum peak values of $\varepsilon_2^y$ and $\varepsilon_2^z$ for $Sb_2S_3$ are around 3.44 $eV$ and 3.48 $eV$, respectively, whereas the maximum values of $\varepsilon_2^y$ and $\varepsilon_2^z$ for $Sb_2Se_3$ are around 2.74 $eV$ and 2.78 $eV$, respectively. Spectral dependences of dielectric functions show the similar features for both materials because the electronic configurations of Se ([Ar],$3d^{10}$ $4s^2$ $4p^2$) and S([Ne], $3s^2$ $3p^3$) are very close to each other. In general, there are various contributions to the dielectric function, but Fig. 4 and Fig. 5 show only the contribution of the electronic polarizability to the dielectric function. The maximum peak values of $\varepsilon_2^y$ and $\varepsilon_2^z$ are in agreement with maximum peak values of theoretical for $Sb_2S_3$ [16]. In the range between 2 $eV$ and 5 $eV$, $\varepsilon_1^z$ decrease with increasing photon-energy, which is characteristics of an anomalous dispersion. In this energy range, the transitions between occupied and unoccupied states mainly occur between S 3p and Se 4p states which can be seen in the DOS displayed in Fig. 3. Furthermore as can be seen from Fig. 4 and Fig. 5, the photon –energy range up to 1.5 $eV$ is characterized by high transparency, no absorption and a small reflectivity. The 1.9-5.0 $eV$ photon energy range is characterized by strong absorption and appreciable reflectivity. The absorption band extending beyond 10 $eV$ up to 15 $eV$ is associated with the transitions from the low-lying valance subband to conduction band. Second, we see that above 12 $eV$, corresponding to the S 3s (Se 4s) and Sb 5p. Also, we remark that the region above 15 $eV$ cannot be interpreted in term of classical oscillators. Above 15 $eV$ $\varepsilon_1$ and $\varepsilon_2$ are dominated by linear features, increasing for $\varepsilon_1$ and decreasing for $\varepsilon_2$.

The corresponding energy-loss functions, $L(\omega)$, are also presented in Fig. 4 and Fig. 5. In this figure, $L_y$ and $L_z$ correspond to the energy-loss functions along the y- and z-directions. The function $L(\omega)$ describes the energy loss of fast electrons traversing the material. The sharp maxima in the energy-loss function are associated with the existence of plasma oscillations [72]. The curves of $L_y$ and $L_z$ in Fig. 4 and Fig. 5 have a maximum near 21.33 and 21.03 $eV$ for $Sb_2S_3$, respectively and 21.90 and 20.16 $eV$ for $Sb_2Se_3$, respectively. These values coincide with the Z point in figure 4 and figure 5. The maximum piks of energy-loss functions are in agreement with maximum peaks of theoretical for $Sb_2S_3$.

The calculated effective number of valence electrons $N_{eff}$ and the effective dielectric constant $\varepsilon_{eff}$ are given in Fig. 6. The effective optical dielectric constant, $\varepsilon_{eff}$, shown in Fig. 6, reaches a saturation value at about 10 $eV$. The photon-energy dependence of $\varepsilon_{eff}$ can be separated into two regions. The first is characterized by a rapid rise and it extends up to 7 $eV$. In the second region the value of $\varepsilon_{eff}$ rises more smoothly and slowly and tends to saturations at the energy 10 $eV$. This means that the greatest contribution to $\varepsilon_{eff}$ arises from interband transitions between 1 $eV$ and 7 $eV$. To determine the contribution made to the static dielectric constant $\varepsilon(0)$ by transitions with frequency $\omega > \omega_0$, we compare the maximum $\varepsilon_{eff}$ with the square of the refractive index ($\varepsilon(0) = n^2$) measured in transparency region Ref. [73]. The difference $\delta\varepsilon_0 = \varepsilon(0) - \varepsilon_{eff}$ ($\delta\varepsilon_0 = 1.16$ for $Sb_2S_3$ and $\delta\varepsilon_0 = 0.52$ for $Sb_2Se_3$) indicates a large contribution of transitions with $\omega > \omega_0$ to the static dielectric constant.

As states above, the $N_{eff}$ determined from the sum rule (Eq. 22) is the effective number of valance electrons per unit cell at the energy $\hbar\omega_0$ (under the condition that all the interband transitions possible at this frequency $\omega_0$ were made). In the case of $Sb_2S_3$ and $Sb_2Se_3$ the value of $N_{eff}$ increases with increasing photon energy and has tendency to saturate near 10 $eV$ and 20 $eV$ (see Fig. 6). Therefore, each of our plots of $N_{eff}$ versus the photon energy for $Sb_2S_3$ and $Sb_2Se_3$ can be arbitrarily divided into two part. The first is characterized by a rapid growth of $N_{eff}$ up to ~8 $eV$ and extend to 12 $eV$. The second part shows a smoother and slower growth of $N_{eff}$ and tends to saturate at energies above 30 $eV$. It is therefore so difficult to choose independent criteria for the estimate of the of valance electrons per unit cell. Recognizing that the two valance subbands are separated from each other and are also separated from the low-lying states of the valance band, we can assume a tendency to saturation at energies such that the transition from the corresponding subbands are exhausted. In other words, since $N_{eff}$ is determined only by the behavior of $\varepsilon_2$ and is the total oscillator strengths, the sections of the $N_{eff}$ curves with the maximum slope, which correspond to the maxima $dN_{eff} / d\hbar\omega$, can be used to discern the appearance of new absorption mechanism with increasing energy (E=5 $eV$, 11.4 $eV$ for $Sb_2S_3$ and E=4.6 $eV$, 12 $eV$ for $Sb_2Se_3$). The values and behavior of $N_{eff}$ and $\varepsilon_{eff}$ for both direction very close to each other.

## Conclusion

In present work, we have made a detailed investigation of the structural, electronic, mechanical, and frequency-dependent linear optical properties of the $Sb_2S_3$ and $Sb_2Se_3$ crystals using the density functional methods. The results of the structural optimization implemented using the LDA are in good agreement with the experimental results. From the present results, we observe that these compounds in mechanically stable. The mechanical properties like shear modulus, Young's modulus, Poisson's ratio, Debye temperature, and shear anisotropic factors are also calculated. Moreover, the ionic contribution to inter atomic bonding for these compounds is dominant. We have revealed that the orthorhombic $Sb_2S_3$ and $Sb_2Se_3$ compounds are in the ground-state configuration and the band structures of these compounds are semiconductor in nature. We have examined photon-energy dependent dielectric functions, some optical properties such as the energy-loss function, the effective number of valance electrons and the effective optical dielectric constant along the y- and z- axes, and mechanical properties. Since there are no experimental elastic data available for $Sb_2S_3$ and $Sb_2Se_3$ compound, we think that the ab initio theoretical estimation is the only reasonable tool for obtaining such important information.

## Tables with captions

**Table 1.** Experimental crystal structure data of orthorhombic $Sb_2S_3$ and $Sb_2Se_3$.

Space group: Pnma—orthorhombic

Atomic positions

| Atom | Wyckoff | x | y | z |
|---|---|---|---|---|
| $Sb_1$ | 4c | 0.5293 | 0.25 | 0.1739 |
| $Sb_2$ | 4c | 0.6495 | 0.75 | 0.4640 |
| $S_1$ | 4c | 0.6251 | 0.75 | 0.0614 |
| $S_2$ | 4c | 0.7079 | 0.25 | 0.3083 |
| $S_3$ | 4c | 0.4503 | 0.75 | 0.3769 |
| | | | | |
| $Sb_1$ | 4c | 0.5304 | 0.25 | 0.1721 |
| $Sb_2$ | 4c | 0.6475 | 0.75 | 0.4604 |
| $Se_1$ | 4c | 0.6289 | 0.75 | 0.0553 |
| $Se_2$ | 4c | 0.7141 | 0.25 | 0.3051 |
| $Se_3$ | 4c | 0.4464 | 0.75 | 0.3713 |

**Table 2.** The calculated equilibrium lattice parameters (a, b, and c), bulk modulus ($B$), and the pressure derivative of bulk modulus ($B'$) together with the theoretical and experimental values for $Sb_2S_3$ and $Sb_2Se_3$ in fractional coordinate.

| Material | Reference | $a$ (Å) | $b$ (Å) | $c$ (Å) | $B(GPa)$ | $B'(GPa)$ |
|---|---|---|---|---|---|---|
| $Sb_2S_3$ | Present (LDA-SIESTA) | 11.29 | 3.83 | 11.20 | 73.64 | 4.42 |
| | Present (LDA-VASP) | 11.02 | 3.81 | 10.79 | | |
| | Theory (EV-GGA) [a] | 11.30 | 3.84 | 11.22 | 71.62 | 5.00 |
| | Experimental[b] | 11.31 | 3.84 | 11.22 | | |
| | Experimental[c] | 11.30 | 3.83 | 11.22 | | |
| | Experimental[d] | 11.27 | 3.84 | 11.29 | | |
| | | | | | | |
| $Sb_2Se_3$ | Present (LDA-SIESTA) | 11.71 | 4.14 | 11.62 | 64.78 | 4.75 |
| | Present (LDA-VASP) | 11.52 | 3.96 | 11.22 | | |
| | Theory (GGA)[e] | 11.91 | 3.98 | 11.70 | | |
| | Experimental[f] | 11.79 | 3.98 | 11.64 | | |
| | Experimental[g] | 11.78 | 3.99 | 11.63 | | |
| | Experimental[h] | 11.77 | 3.96 | 11.62 | | |

[a]Reference [16]

[b]Reference [32]

[c]Reference [33]

[d]Reference [34]

[e]Reference [17]

[f]Reference [2]

[g]Reference [35]

[h]Reference [36]

**Table 3.** The calculated elastic constants (in GPa) for $Sb_2S_3$ and $Sb_2Se_3$.

| Material | Reference | $C_{11}$ | $C_{22}$ | $C_{33}$ | $C_{12}$ | $C_{13}$ | $C_{23}$ | $C_{44}$ | $C_{55}$ | $C_{66}$ |
|---|---|---|---|---|---|---|---|---|---|---|
| $Sb_2S_3$ | Present (LDA-SIESTA) | 133.56 | 115.64 | 121.50 | 38.21 | 45.98 | 69.14 | 84.74 | 58.93 | 41.01 |
| | Present (LDA-VASP) | 134.41 | 139.29 | 118.45 | 38.20 | 57.37 | 69.99 | 73.72 | 59.39 | 41.72 |
| $Sb_2Se_3$ | Present (LDA-SIESTA) | 101.56 | 89.94 | 84.60 | 34.13 | 43.66 | 48.58 | 54.92 | 40.84 | 30.37 |
| | Present (LDA-VASP) | 118.88 | 118.36 | 105.62 | 33.07 | 53.12 | 60.16 | 65.28 | 54.67 | 36.19 |

**Table 4.** The calculated isotropic bulk modulus (B, in GPa), shear modulus (G, in GPa), Young's modulus (E, in GPa) and Poisson's ratio for $Sb_2S_3$ and $Sb_2Se_3$ compounds.

| Material | Reference | $B_R$ | $B_V$ | $B$ | $G_R$ | $G_V$ | $G$ | $E$ | $\upsilon$ |
|---|---|---|---|---|---|---|---|---|---|
| $Sb_2S_3$ | Present (LDA-SIESTA) | 74.93 | 75.26 | 75.10 | 43.55 | 51.43 | 47.49 | 117.66 | 0.24 |
| | Present (LDA-VASP) | 80.14 | 80.36 | 80.25 | 44.14 | 50.07 | 47.11 | 118.20 | 0.25 |
| $Sb_2Se_3$ | Present (LDA-SIESTA) | 58.72 | 58.76 | 58.74 | 30.89 | 35.21 | 33.05 | 83.49 | 0.26 |
| | Present (LDA-VASP) | 70.42 | 70.62 | 70.52 | 38.75 | 44.32 | 41.54 | 104.52 | 0.25 |

**Table 5.** The shear anisotropic factors $A_1$, $A_2$, $A_3$, and $A_{comp(\%)}$, $A_{shear(\%)}$.

| Material | Reference | $A_1$ | $A_2$ | $A_3$ | $A_{comp(\%)}$ | $A_{shear}(\%)$ |
|---|---|---|---|---|---|---|
| $Sb_2S_3$ | Present (LDA-SIESTA) | 2.09 | 2.38 | 0.95 | 0.22 | 8.29 |
| | Present (LDA-VASP) | 2.14 | 2.50 | 1.20 | 0.14 | 6.29 |
| $Sb_2Se_3$ | Present (LDA-SIESTA) | 2.22 | 2.11 | 0.99 | 0.03 | 6.50 |
| | Present (LDA-VASP) | 2.21 | 2.11 | 0.85 | 0.14 | 6.71 |

**Table 6.** The density, longitudinal, transverse, average elastic wave velocities, and hardness together with the Debye temperature for $Sb_2S_3$ and $Sb_2Se_3$.

| Material | Reference | $\rho(g/cm^3)$ | $v_l\,(m/s)$ | $v_t\,(m/s)$ | $v_m\,(m/s)$ | $\theta_D\,(K)$ | Hv (GPa) |
|---|---|---|---|---|---|---|---|
| $Sb_2S_3$ | Present (LDA-SIESTA) | 4.65 | 5455.88 | 3195.76 | 3543.32 | 364.41 | 8.14 |
| | Present (LDA-VASP) | 4.98 | 5570.75 | 3196.73 | 3550.86 | 364.14 | 7.27 |
| | Experimental[b] | 4.61 | | | | | |
| $Sb_2Se_3$ | Present (LDA-SIESTA) | 5.67 | 4258.13 | 2414.32 | 2684.51 | 262.78 | 4.85 |
| | Present (LDA-VASP) | 6.23 | 4627.39 | 2657.94 | 2952.13 | 292.50 | 6.54 |
| | Experimental[b] | 5.88 | | | | | |

**Table 7.** Energy band gap for $Sb_2S_3$ and $Sb_2Se_3$, obtained from SIESTA.

| Material | Reference | $E_g(eV)$ |
|---|---|---|
| $Sb_2S_3$ | Present | 1.18 direct |
| | Experimental[a] | 1.78 indirect-2.25 direct |
| | Experimental[b] | 1.63 indirect-1.72 direct |
| | Experimental[c] | 1.56 direct |
| | Experimental[d] | 1.64 |
| | Experimental[e] | 1.71 |
| | Experimental[f] | 2.2 (300 K)-1.60 (473 K) direct |
| | Theory[g] | 1.55 |
| | Theory[h] | 1.76 |
| $Sb_2Se_3$ | Present | 0.99 indirect-1.07 direct |
| | Experimental[ı] | 1.0 direct |
| | Experimental[i] | 1.5 direct |
| | Experimental[j] | 1.1 indirect |
| | Experimental[k] | 1.0-1.2 indirect |
| | Theory[g] | 1.14 |

[a]Reference [60]

[b]Reference [61]

[c]Reference [62]

[d]Reference [63]

[e]Reference [64]

[f]Reference [65]

[g]Reference [15]

[h]Reference [16]

[ı]Reference [66]

[i]Reference [67]

[j]Reference [62]

[k]Reference [68]

## Figure captions

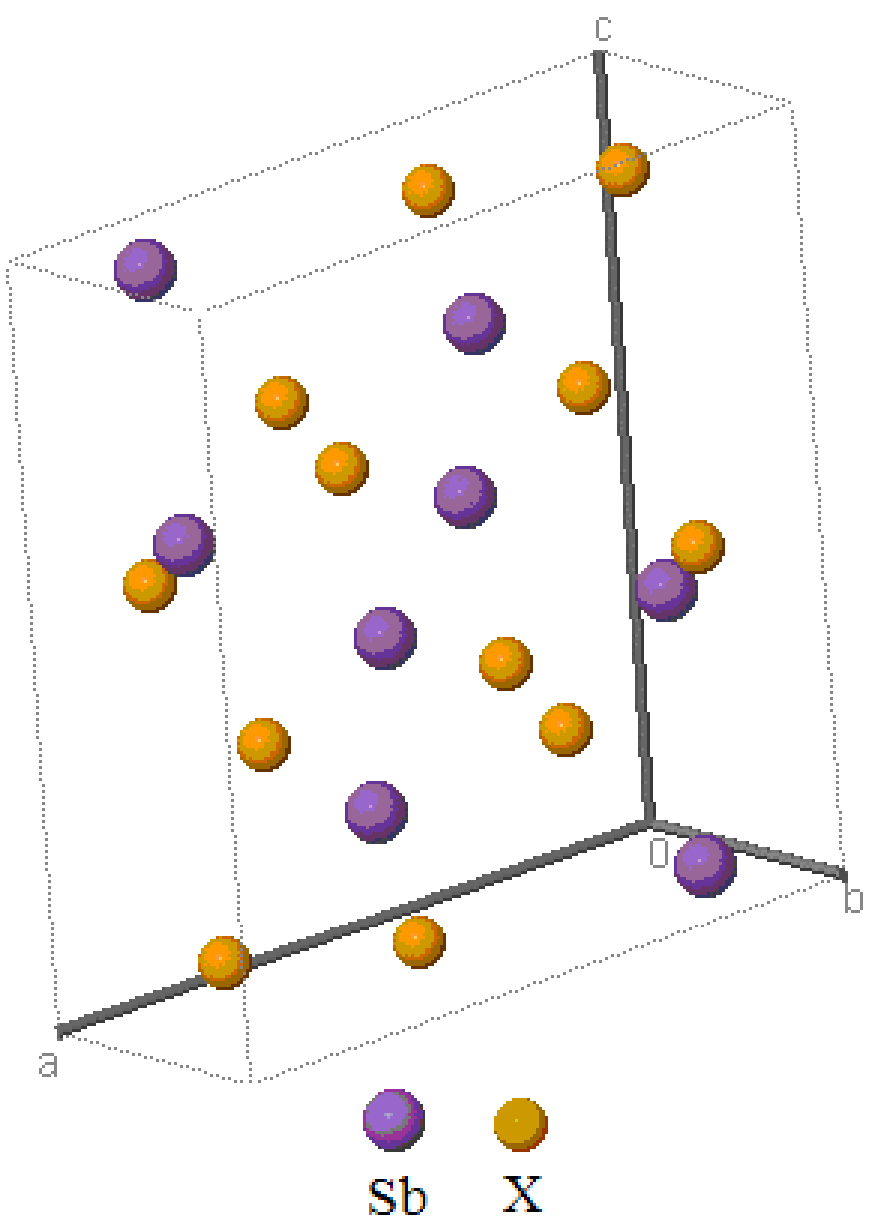

**Fig 1.** Crystal structure of $Sb_2X_3$ (X=S, Se).

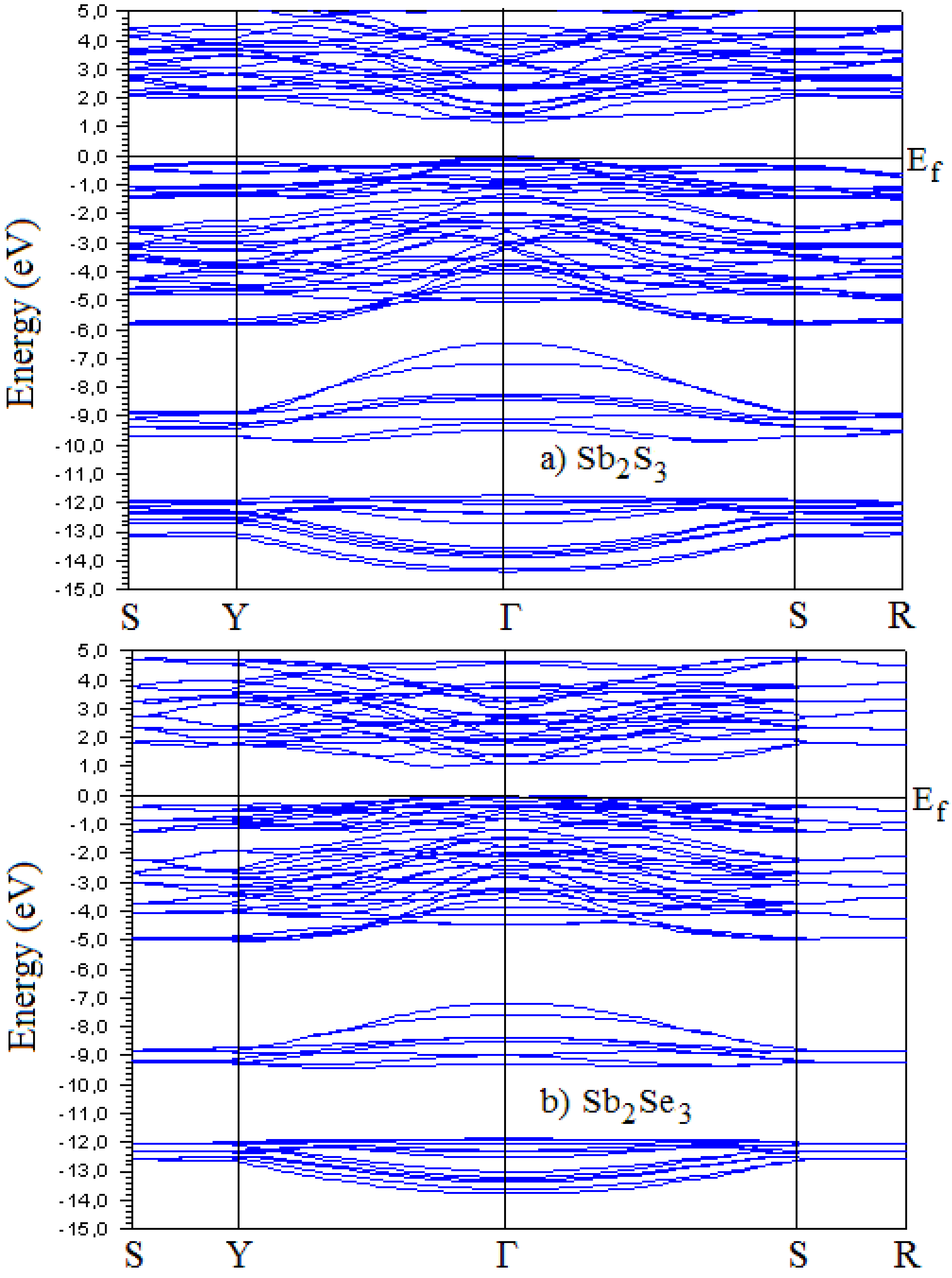

**Fig 2.** Energy band structure for $Sb_2S_3$ and $Sb_2Se_3$.

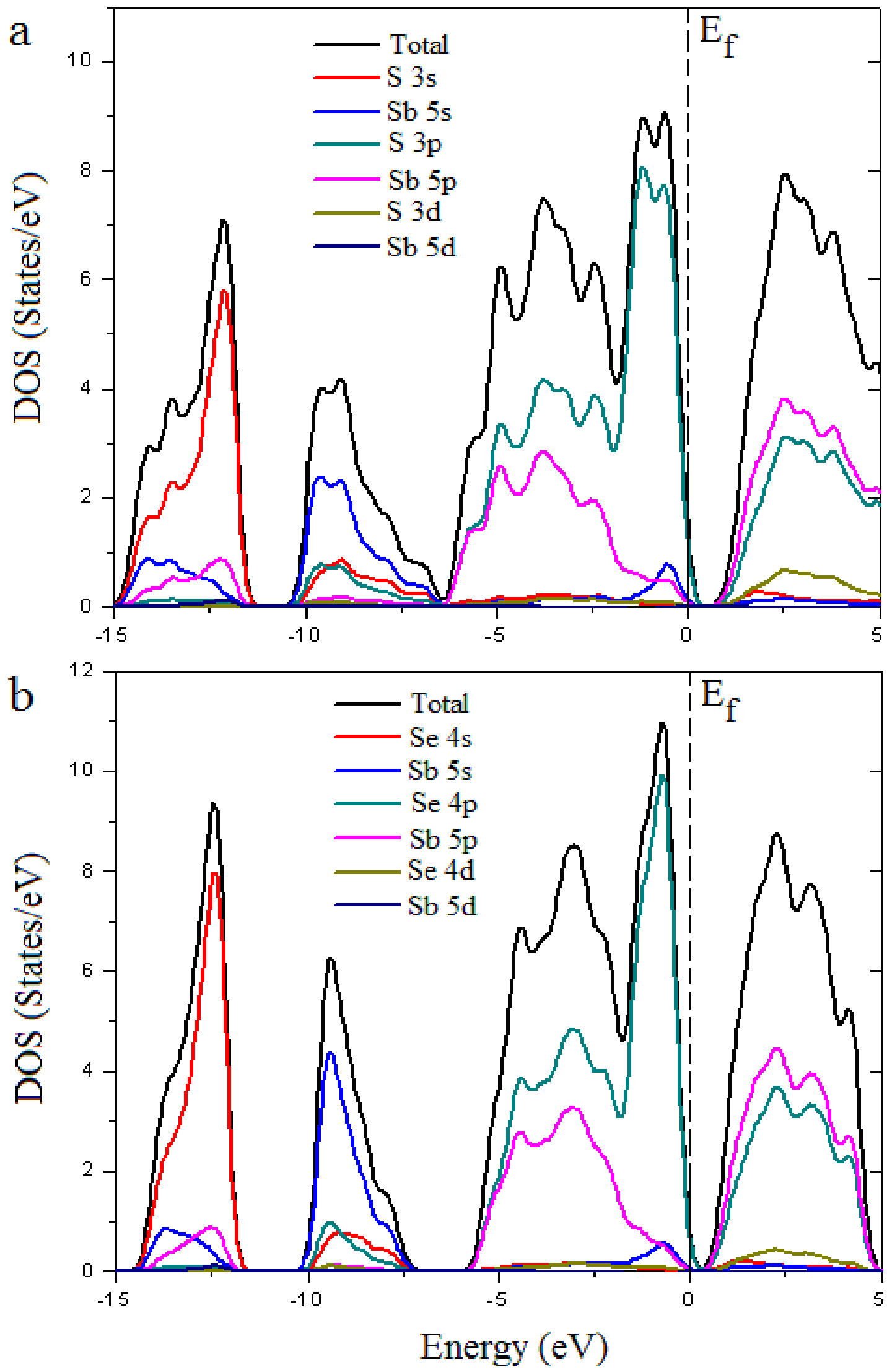

**Fig 3.** The total (DOS) and projected density of states for a) $Sb_2S_3$ and b) $Sb_2Se_3$.

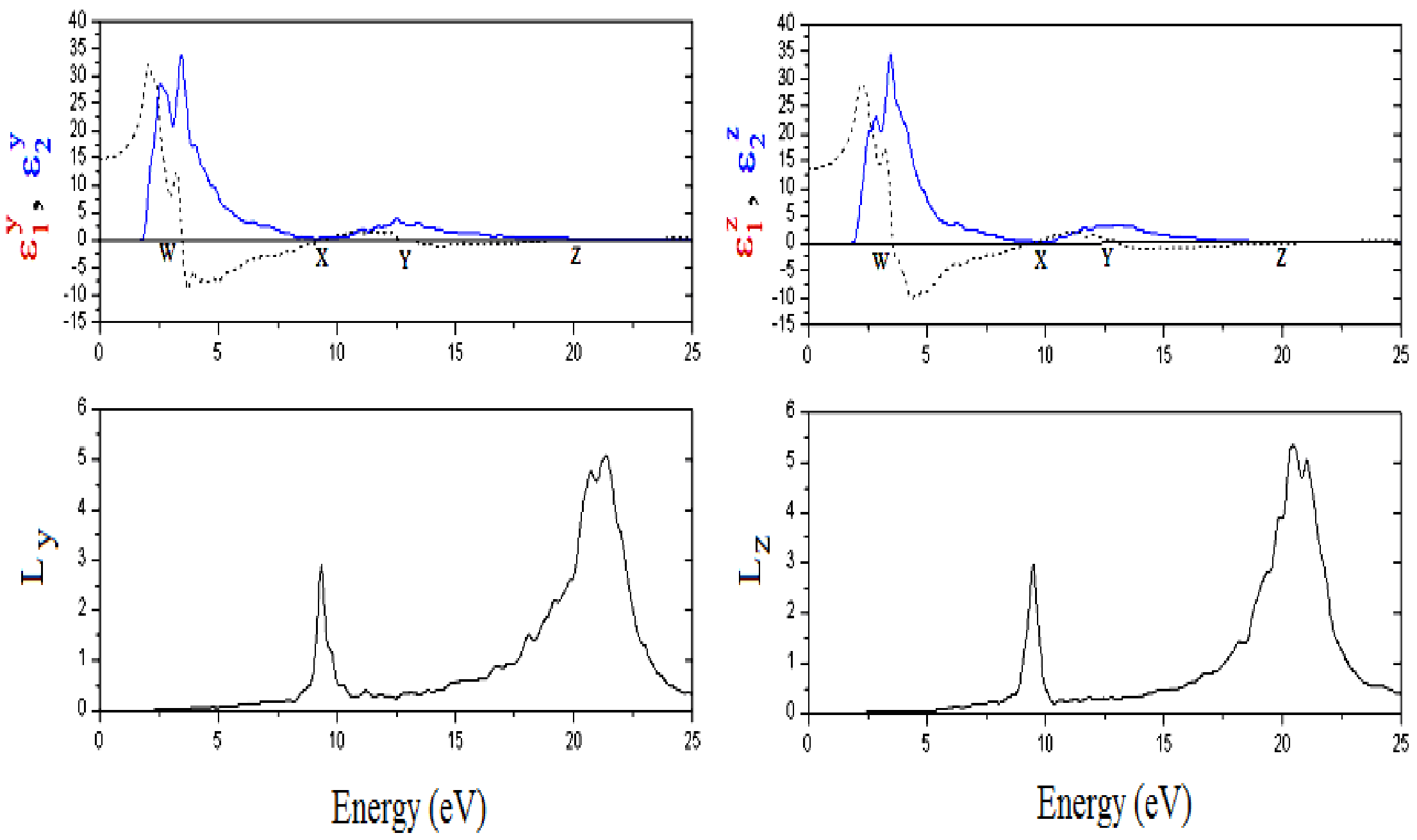

**Fig 4.** Energy spectra of dielectric function $\varepsilon = \varepsilon_1 - i\varepsilon_2$ and energy-loss function (L) along the y- and z-axes for $Sb_2S_3$

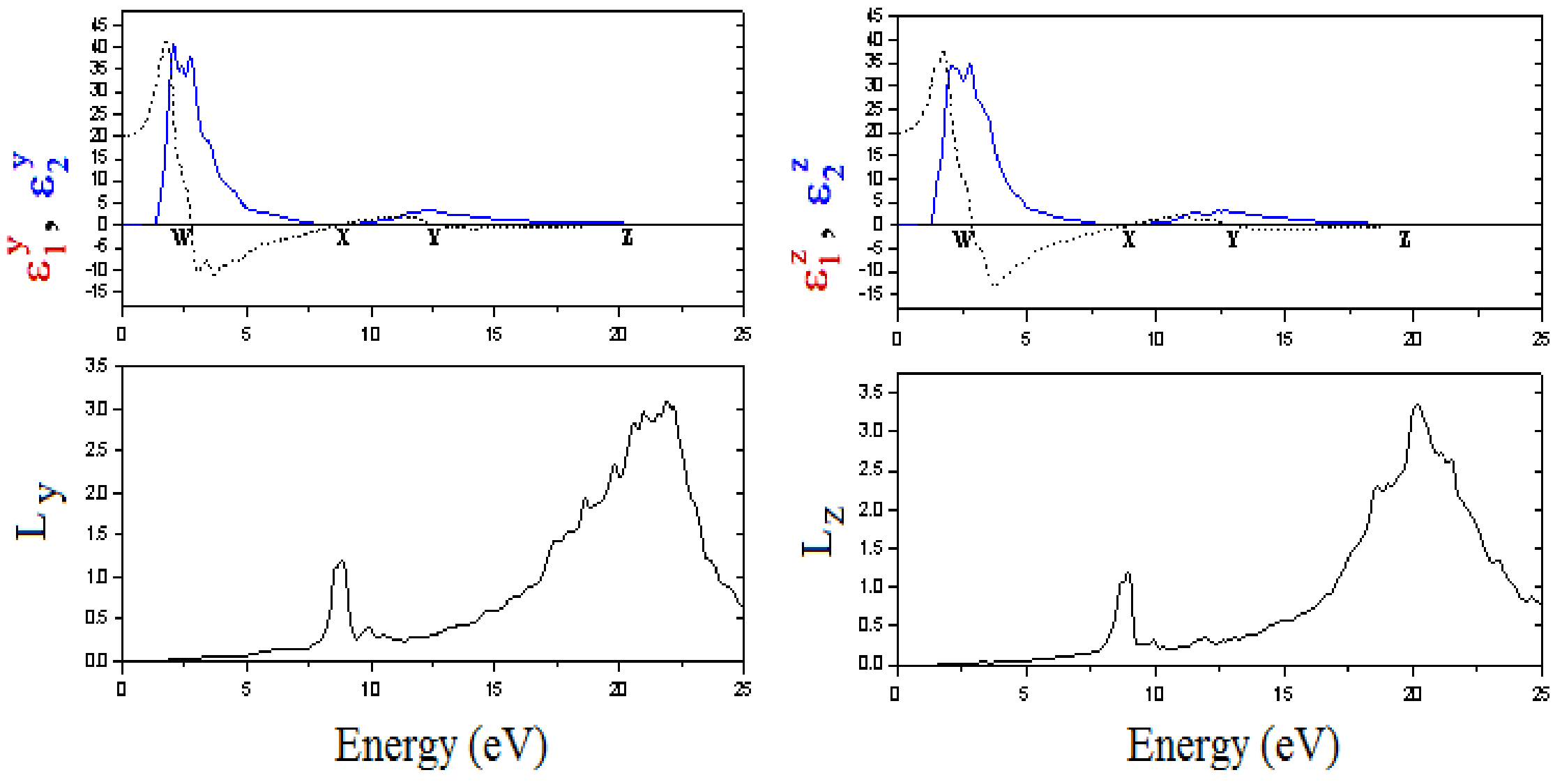

**Fig 5.** Energy spectra of dielectric function $\varepsilon = \varepsilon_1 - i\varepsilon_2$ and energy-loss function (L) along the y- and z-axes for $Sb_2Se_3$

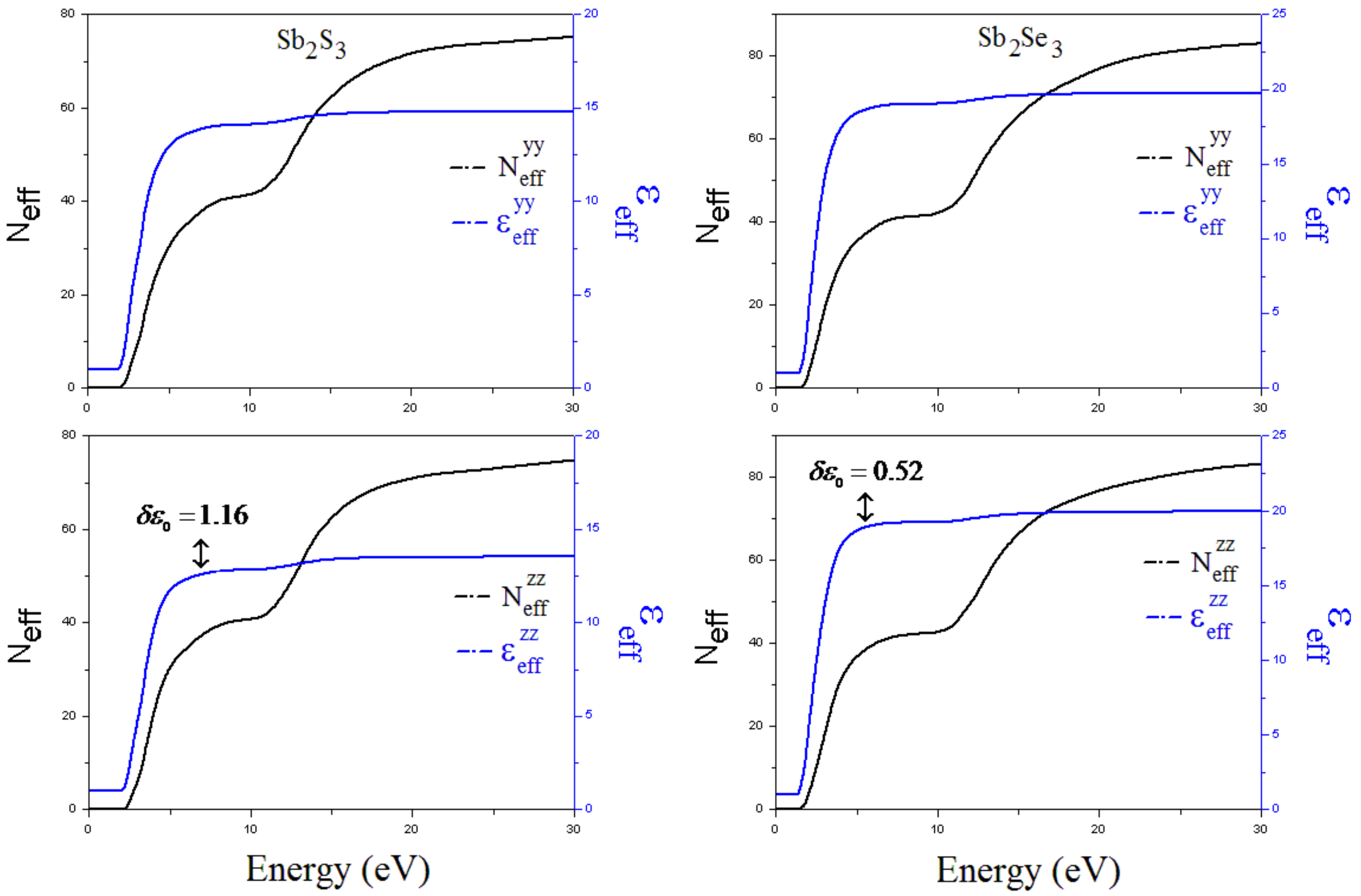

**Fig 6.** Energy spectra of $N_{eff}$ and $\varepsilon_{eff}$ along the y- and z- axes